# Can thermal nonreciprocity improve the radiative cooling efficiency?

Mengqi Liu[1,2], Shenghao Jin[1], Chenglong Zhou[3], Boxiang Wang[4,5,6], Changying Zhao[1]*, Cheng-Wei Qiu[2]*

[1]*Institute of Engineering Thermophysics, MOE Key Laboratory for Power Machinery and Engineering, School of Mechanical Engineering, Shanghai Jiao Tong University, Shanghai 200240, China*

[2]*Department of Electrical and Computer Engineering, National University of Singapore, Singapore, Singapore*

[3]*School of Energy Science and Engineering, Harbin Institute of Technology, Harbin 150001, China*

[4]*2020 X-Lab, Shanghai Institute of Microsystem and Information Technology, Chinese Academy of Sciences, Shanghai 200050, China*

[5]*State Key Laboratory of Transducer Technology, Shanghai Institute of Microsystem and Information Technology, Chinese Academy of Sciences, Shanghai 200050, China*

[6]*School of Graduate Study, University of Chinese Academy of Sciences, Beijing 100049, China*

* *Email: changying.zhao@sjtu.edu.cn; chengwei.qiu@nus.edu.sg*

## Abstract

Probably not.

**Main text:**

Recently, radiative cooling (RC) [1–3] as a promising cooling technology to harvest the coldness of the universe **(Fig. 1A)** has been widely studied, which has attracted broad interest in both fundamental sciences and practical applications, ranging from passive cooling of buildings or solar cells, dew harvesting, renewable energy harvesting, outdoor personal thermal management, etc. The RC technology aims to dissipate thermal radiation from Earth to outer space through the waveband of 8-13μm known as the atmospheric window (AW). Thereby, much attention has been paid to engineering spectral emissivity of thermal emitters (i.e., selective thermal emitters (SEs) in **Fig. 1B**) using various materials and nanostructures, including multilayer, thin films containing randomly distributed nanoparticles, porous structures, etc. To enhance the cooling performance including the cooling power $P_{cool}$ (W/m$^2$) and cooling temperature $\Delta T$ (°C or K), we have witnessed remarkable progress in optimizing the coolers' materials, structures, angle responses, etc. RC can achieve cooling power $>100\text{W/m}^2$ at ambient temperature $T_{amb}$ and cool down a sky-facing object to more than ~50°C below $T_{amb}$ [4]. Such impressive cooling performance is only available in sunny, dry, cloudless surroundings, then it could be easily destroyed when the weather conditions become bad [5]. Besides, the reported RC devices, thus far, are always constrained by a certain theoretical limit [6], as compared in **Fig. 1C** and Table S1. The modified cooling power $\tilde{P}_{cool}(T_{amb}) = P_{cool} + P_{sun}$ is employed to eliminate the difference in solar power absorbed in different works. The theoretical limit of the broadband perfect emitter (BE in **Fig. 1B**) marked in the black line is calculated under the clearest standard weather condition (ATRAN, Fig. S5) with broad and high atmospheric transmission. All the existing radiative coolers underperform below the theoretical bound. Such behaviors inspire us to think of the possibility of breaking this fundamental limit and further improving the cooling efficiency even in hazy, humid, cloudy weather conditions.

According to energy conversion and thermodynamics of RC, the modified cooling power of a sky-facing thermal emitter can be calculated by [3] $\tilde{P}_{cool}(T) = P_{cool} + P_{sun} = P_{emi}(T) - P_{amb}(T_{amb}) - P_c$, where $P_c = h(T_{amb} - T)$ is the heat load on the emitter because of the conductive and convective heat exchange with the surroundings. The $P_{emi} = \int d\Omega \cos\theta \int_0^\infty d\lambda I_{BB}(T,\lambda) e(\lambda,\theta)$ is the power emitted from the thermal emitter, in which $\Omega$ is a solid angle, $\theta$ is the angle between the direction of the solid angle and the normal direction of the surface. The $P_{amb}(T_{amb}) = \int d\Omega \cos\theta \int_0^\infty d\lambda I_{BB}(T_{amb},\lambda) e_{amb}(\lambda,\theta)\alpha(\lambda,\theta)$ shows the portion of downward radiation absorbed by the emitter from the atmosphere, with $e_{amb}(\lambda,\theta) = 1 - t(\lambda)^{1/\cos\theta}$ being the emissivity of the atmosphere ($t(\lambda)$ is the transmission coefficient of the atmosphere in the zenith direction). The $I_{BB}(T,\lambda)$ denotes spectral irradiance of a blackbody. The $\varepsilon(\lambda,\theta)$ and $\alpha(\lambda,\theta)$ are the emitter's emissivity and absorptivity at a wavelength $\lambda$ and angle $\theta$. **Figure 1D** gives the spectral irradiance under a condition of mid-latitude winter (MLW) atmosphere (light blue area in **Fig. 1B**) between $E_{emi}(\lambda,T) = I_{BB}(\lambda,T_{amb})\int d\Omega \cos\theta$ and $E_{amb}(\lambda,T_{amb}) = I_{BB}(T_{amb},\lambda)\int e_{amb}(\lambda,\theta)\alpha(\lambda,\theta) d\Omega \cos\theta$ at $T_{amb} = 20^\circ C$. Here we choose the MLW atmosphere as an example since it shows only one typical AW at 8-13μm. Generally, the value of $P_{amb}(T_{amb})$ is calculated based on $\alpha(\lambda,\theta) = e(\lambda,\theta)$ rooted in Kirchhoff's law of thermal radiation [7]. The maximum net modified cooling power of BE and SE give $\tilde{P}_{cool,BE}^{max} = 98.7 W/m^2$ (green area) and $\tilde{P}_{cool,SE}^{max} = 89.2 W/m^2$, respectively. The conventional wisdom for radiative cooling is to ensure high emissivity through atmospheric transparent windows, but the atmospheric absorption (white area) actually becomes responsible for a large portion of parasitic loss and imposes an upper bound of the cooling power limit. So, it is necessary to consider mitigating the adverse impact of non-atmospheric windows (non-AWs).

Thermal nonreciprocity [8], which introduces Lorentz nonreciprocity in thermal sciences, may provide an alternative solution. By introducing nonreciprocal strategies like magneto-optical (MO) materials, time-space modulation, or optical nonlinear effects in thermal nanostructures, it is possible to break the equality of spectral/angular emissivity $e(\lambda,\theta)$ and absorptivity $\alpha(\lambda,\theta)$, providing a new avenue to mitigate the loss mechanism caused by Kirchhoff's law and improve the energy converse/transfer efficiency. Recent experimental works have already demonstrated broadband nonreciprocal thermal radiation [9] using gradient-doped MO semiconductors layers in the mid-infrared region under a moderate external magnetic field **B**. By utilizing the nonreciprocal thermal emitters/absorbers in energy devices, the ultimate efficiency (i.e., Landsberg limit 93.3%) of solar energy harvesting can be reached at the thermodynamic limit [10]. Similar Landsberg schemes have also been generalized to nonreciprocal thermophotovoltaics systems [11,12], showing great potential in next-generation nonreciprocal energy devices. Thereby, one question naturally arises: ***Can thermal nonreciprocity help to further improve the radiative cooling efficiency? Our counterintuitive answer is "Probably not".***

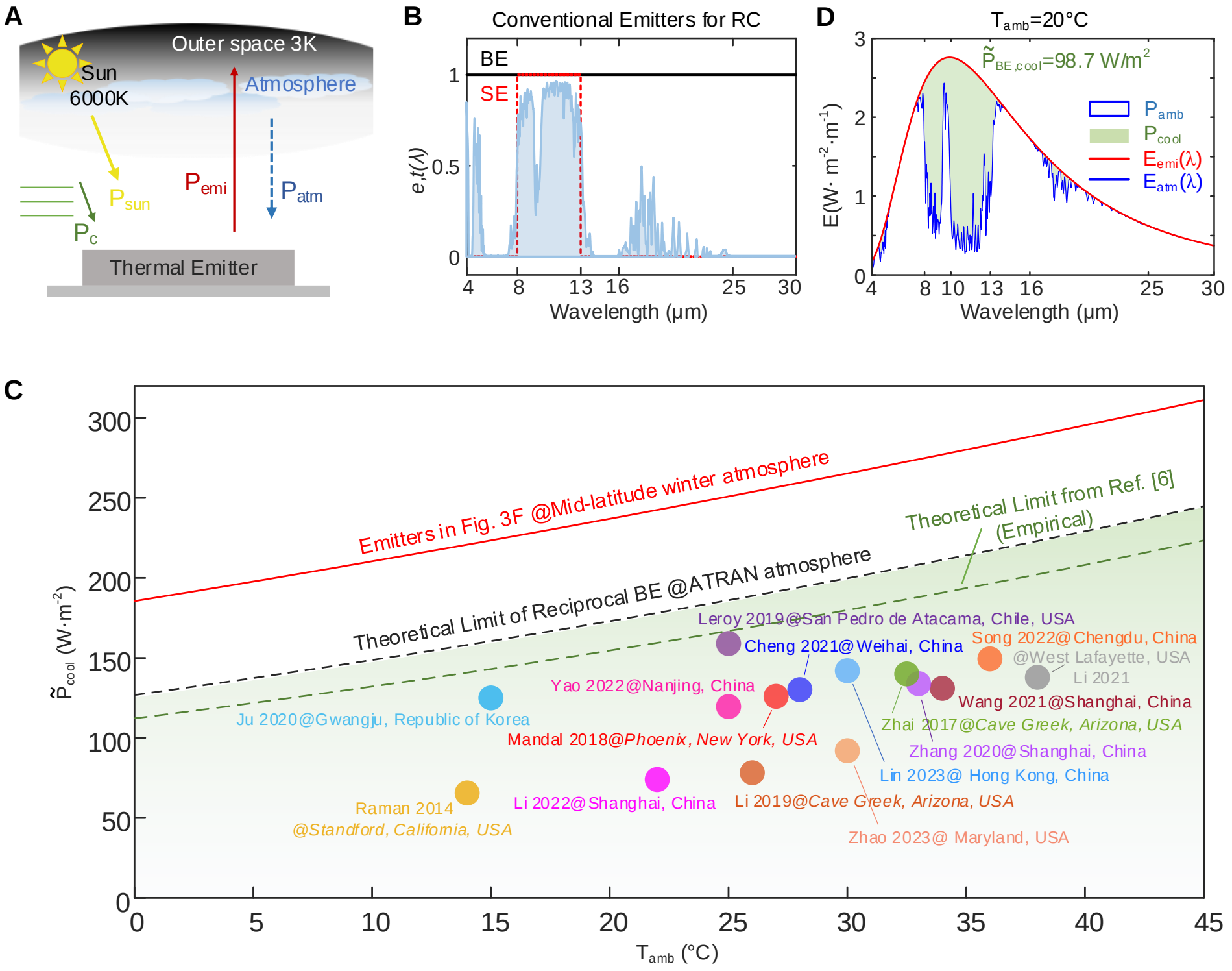

**Figure 1 Basic principles and limitations of conventional RC.** (A) Radiative heat-exchange process of day-time radiative cooling. (B) Two commonly used thermal emitters for conventional RC. The shadowed blue area shows the atmospheric transparency under MLW conditions. (C) Comparison of the modified cooling power $\tilde{P}_{cool}$ reported in existing work listed in Table S1, along with the theoretical limits (dashed lines for reciprocal and full line for nonreciprocal cases). (D) Spectral irradiance (red line) and parasitic absorption loss (blue line) of the BE.

Here, we propose that nonreciprocal RC may be not as efficient as reciprocal counterparts in some cases. By classifying the angular and spectral dependence of general thermal emitters, we emphasize that the realization of efficient nonreciprocal RC strictly requires nonreciprocal and asymmetric selective thermal emitters with high emissivity through AWs and low absorptivity through non-AWs, along with a proof-of-concept demonstration of the possibility of breaking the reciprocal cooling efficiency. The important roles of nonreciprocity, geometric asymmetry, and non-AWs will be discussed, followed by the prospects and challenges in nonreciprocal RC.

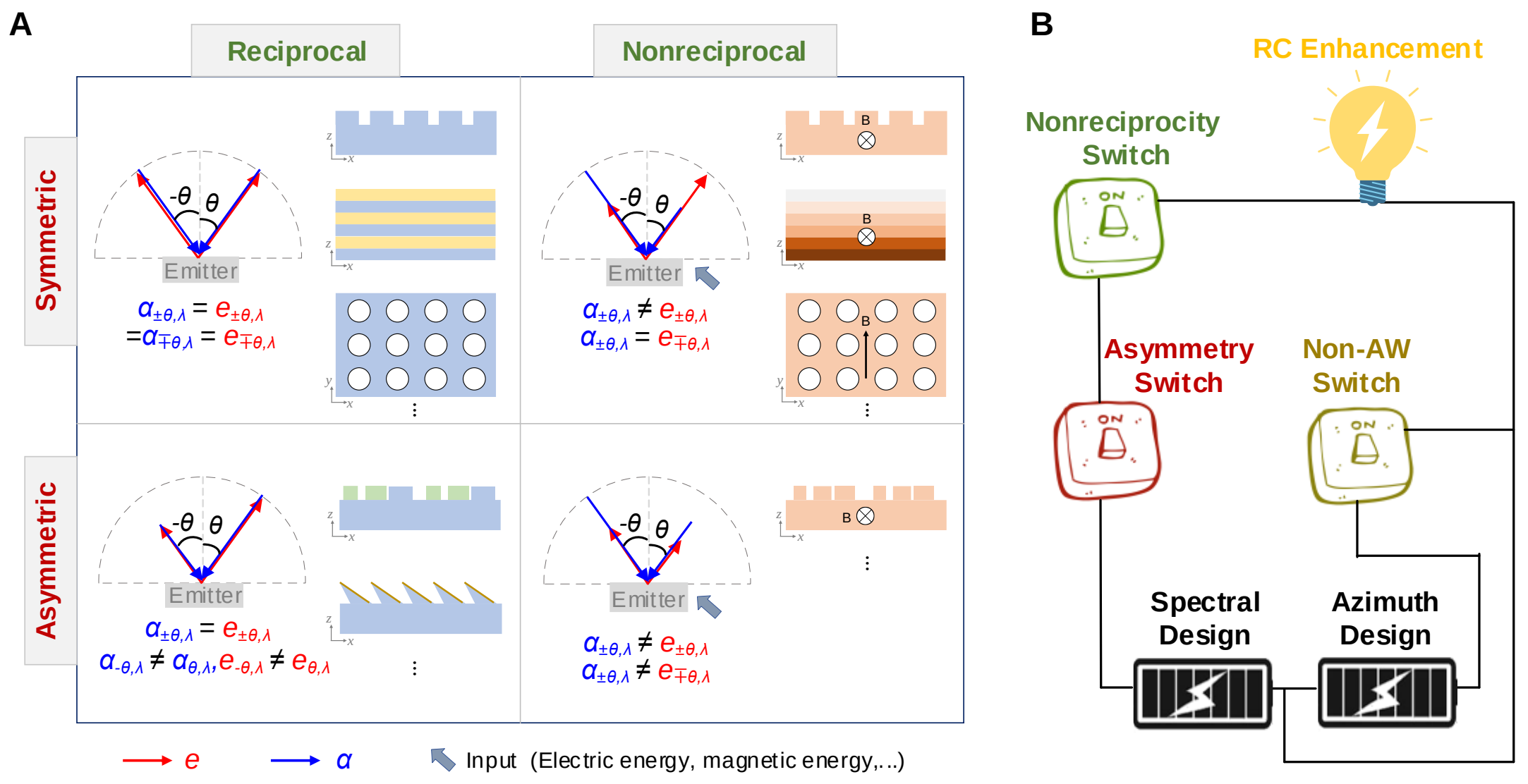

**Figure 2 The role of thermal nonreciprocity and geometric asymmetry in designing RC thermal emitters.** (A) Relationship of spectral/angular emissivity and absorptivity for thermal emitters classified by reciprocity and geometric symmetry, along with the corresponding structures. (B) Schematic of the functions of nonreciprocity, asymmetry, and non-AW in improving the radiative cooling efficiency.

**Thermal nonreciprocity and geometric asymmetry in designing RC thermal emitters.**

In nonreciprocal systems, the spectral/angular responses of emissivity and absorptivity over the symmetric hemisphere will not be consistent with each other. Then, the emission power and absorption power of the sky-facing emitter should be modified to

$$P_{emi}(T)=\int_0^{\pi}d\varphi\int_0^{\frac{\pi}{2}}d\theta\sin\theta\cos\theta\int_0^{\infty}d\lambda I_{BB}(T,\lambda)\left[e(\lambda,\theta)+e(\lambda,-\theta)\right] \tag{1}$$

$$P_{amb}(T_{amb})=\int_0^{\pi}d\varphi\int_0^{\frac{\pi}{2}}d\theta\sin\theta\cos\theta\int_0^{\infty}d\lambda I_{BB}(T_{amb},\lambda)e_{amb}(\lambda,\theta)\left[\alpha(\lambda,\theta)+\alpha(\lambda,-\theta)\right] \tag{2}$$

Thus, the modified net cooling power is directly determined by the difference of $\tilde{P}_{cool}(T=T_{amb})\propto\Delta_{non}=\left[e(\lambda,\theta)+e(\lambda,-\theta)\right]-e_{amb}\left[\alpha(\lambda,\theta)+\alpha(\lambda,-\theta)\right],\theta\in\left[0,\pi/2\right]$. For reciprocal cases, this equation reduces to $\Delta_{rec}=2e(\lambda,\theta)\left[1-e_{amb}(\lambda,\theta)\right],\theta\in\left[0,\pi/2\right]$, being consistent with the analysis in Ref. [6]. The following important question is what is the clear relationship between $e(\lambda,\pm\theta)$ and $\alpha(\lambda,\pm\theta)$ in a general system? In theory, both nonreciprocity and geometric symmetry play essential roles. As summarized in **Figs. 2A**, in terms of commonly studied reciprocal thermal emitters obeying Kirchhoff's law, there is $\alpha(\lambda,\pm\theta)=e(\lambda,\pm\theta)=\alpha(\lambda,\mp\theta)=e(\lambda,\mp\theta)$ for symmetric structures with either inversion symmetry or translational symmetry like multilayer and periodical metasurfaces (inset). When the symmetry of structures is broken, one can observe reciprocal but asymmetric absorption or emission spectra $\alpha(\lambda,\pm\theta)=e(\lambda,\pm\theta),\alpha(\lambda,\theta)\neq\alpha(\lambda,-\theta),e(\lambda,\theta)\neq e(\lambda,-\theta)$ , which has been experimentally demonstrated using asymmetric gratings [13] or micro-wedge geometry [14]. For nonreciprocal systems usually with input external energy like magnetic energy or electric energy to break Lorentz reciprocity, the relationship between spectral/angular emissivity and absorptivity will be complex. For example, as for MO systems with an applied **B** widely studied in nonreciprocal thermal photonics [8], it is known that the relation $\alpha(\lambda,\pm\theta,\mathbf{B})=e(\lambda,\pm\theta,\mathbf{B})$ will not hold whether the structure is symmetric or not. But the geometric symmetries still lead to $\alpha(\lambda,\pm\theta,\mathbf{B})=e(\lambda,\mp\theta,\mathbf{B})$. When thermal emitters are neither reciprocal nor geometric symmetric [15], one can expect no specific relation between $\alpha(\lambda,\pm\theta,\mathbf{B})$ and $e(\lambda,\pm\theta,\mathbf{B})$. Even so, due to Onsager-Casimir relations, the relation of $\alpha(\lambda,\pm\theta,\mathbf{B})=e(\lambda,\pm\theta,-\mathbf{B})$ will always hold in two magnetized systems with opposite directions of external magnetic fields.

On the other hand, as discussed before, the atmospheric condition $e_{amb}(\lambda,\theta)$ also directly influences the value of $\triangle_{\text{non}}$. It is known that, for conventional RC, it is necessary to improve the emissivity within the atmospheric window to efficiently emit radiation power to outer space. However, nonreciprocal thermal emitters could be a gamer changer, where the non-AW will play an essential role. **Figure 2B** illustrates the importance of *nonreciprocity* (green switch), *geometric asymmetry* (red switch), and *non-AW* (yellow switch) in cooling power enhancement. Analogically, it is possible to lighten the "*RC Enhancement*" bulb compared with reciprocal cases when turning on both nonreciprocity and asymmetry switches (azimuth design). In addition, the spectral design of nonreciprocal thermal emitters can further improve cooling efficiency by turning on the non-AW switch. Next, we will give a detailed analysis of how these three factors affect the cooling performance.

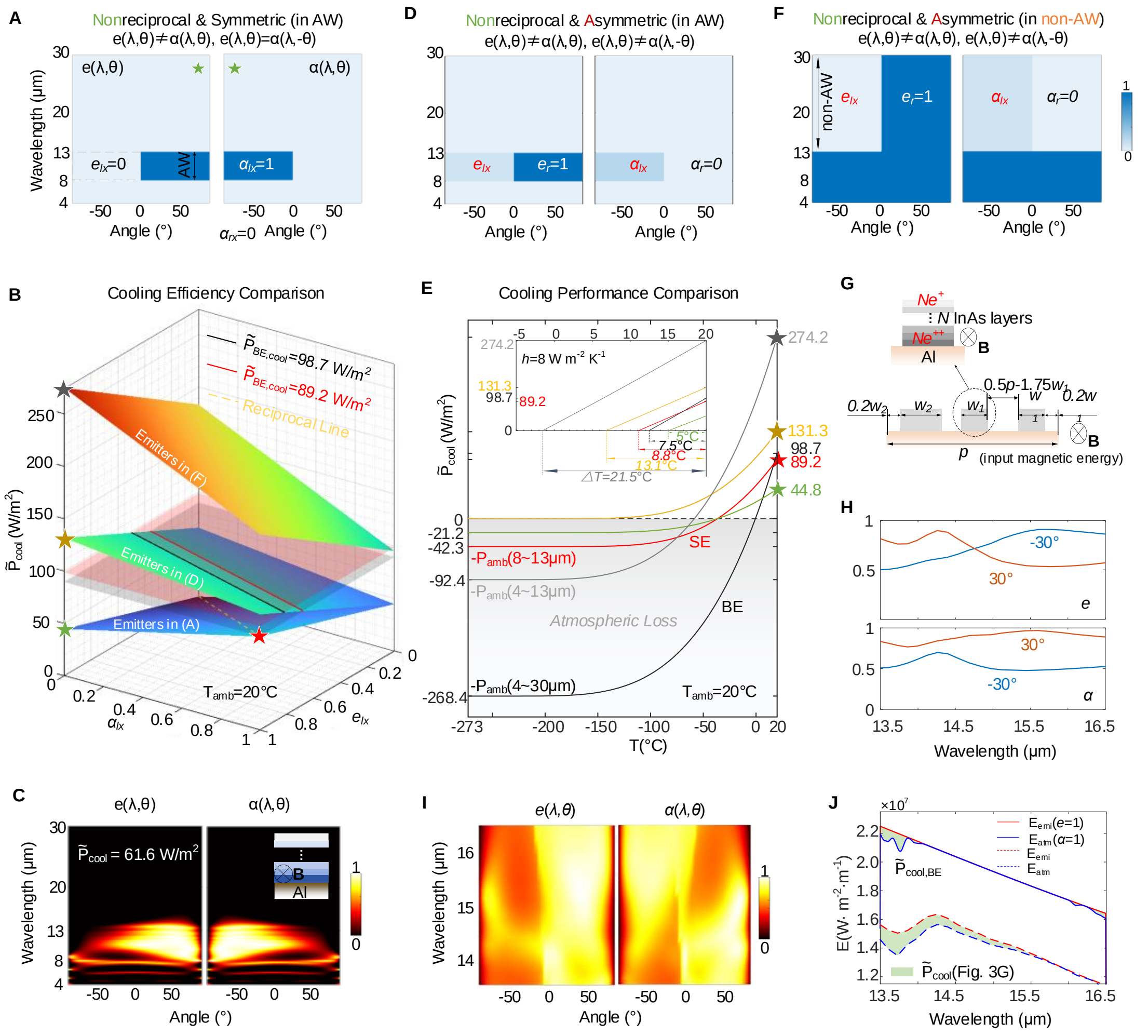

**Figure 3 Discussing the role of nonreciprocity, geometric asymmetry and non-AW in cooling performance enhancement.** (A) Nonreciprocal emission/absorption spectra with geometric symmetry within the AW waveband, along with the $\tilde{P}_{cool}$ (bottom surface) as functions of $e_{lx}$ and $\alpha_{lx}$ in (B). Red star denotes the case of reciprocal SE, and green star denotes the typical case with $e_{lx}=1$ and $\alpha_{lx}=0$. (C) Angle-resolved emission and absorption spectra for the MO multilayer structure (inset) (B=5T). (D) Nonreciprocal emission/absorption spectra without geometric symmetry within the AW waveband $e_r=1, \alpha_r=0$, along with (B) the results of $\tilde{P}_{cool}$ (middle surface) as functions of $e_{lx}, \alpha_{lx}$ in (D). (E) The theoretical limit of the radiative cooling power $\tilde{P}_{cool}(T)$ of different thermal emitters marked in (B) without conductive and convective heat exchange (inset with $h=8\mathrm{W/(m^2\cdot K)}$). (F) Nonreciprocal emission/absorption spectra without geometric symmetry within the non-AW waveband with $e(8-13\mu m,-\pi/2\sim\pi/2)=\alpha(8-13\mu m,-\pi/2\sim\pi/2)=1$, $e(13-30\mu m,0\sim\pi/2)=1$ and $\alpha(13-30\mu m,0\sim\pi/2)=0$. (G) Feasible candidate of nonreciprocal thermal emitters for nonreciprocal radiative cooling. (I) Asymmetric and nonreciprocal emission and absorption spectra, along with (H) spectral at typical angles $\theta=\pm30^\circ$. (J) Compared spectral irradiance (red) and atmospheric absorption (blue line) for ideal BE (full lines) and proposed asymmetric nanostructure in (G). The shadowed green areas are the available cooling power for two cases in this non-AW waveband.

**Merely breaking thermal reciprocity will weaken RC performance.**

The general idea for radiative cooling is to design thermal emitters with high emissivity (absorptivity) through AW. We first consider a nonreciprocal SE in **Figs. 3A** with no angular and spectral dispersion but geometric symmetry. Due to geometric symmetry constraints, there must be $\alpha(\lambda,\text{-}\theta)=e(\lambda,\theta), e(\lambda,\text{-}\theta)=\alpha(\lambda,\theta),\ \ \theta\in[0,\pi/2]$ , but the value of $e_{lx}=e(\lambda,-\theta)$ and $\alpha_{lx}=\alpha(\lambda,-\theta)$ can be independently controlled. **Figure 3B** gives the results of $\tilde{P}_{cool}(T=T_{amb}\text{=}20^{\circ}\text{C})$ under MLW condition as $e_{rx}$ and $\alpha_{rx}$ .change (bottom surface). The maximum cooling power (red star) can only be achieved in the reciprocal case with $\tilde{P}_{cool,SE}^{\max}=89.2\text{W/m}^2$, which is superior to all the nonreciprocal cases considered in **Fig. 3A**. For example, in the case of $e_{rx}=1,\alpha_{rx}=0$ (green star in **Fig. 3B**), the available net cooling power is limited to $44.8\text{W/m}^2$. To be honest, up to now, the majority of proposed nonreciprocal thermal emitters belong to this symmetric type. Here we give an example of a nonreciprocal multilayer structure (inset in **Fig. 3C**), which is composed of thirteen InAs layers (200-nm thickness for each layer) with doping concentrations $N_e$ from $2.5\times10^{18}\text{cm}^{-3}$ (top) to $7.3\times10^{18}\text{cm}^{-3}$ (bottom) with an interval $\Delta Ne=0.4\times10^{18}\text{cm}^{-3}$ sitting on Al substrate. The details of used materials with permittivity tensor are given in supplementary note 2. The angle-resolved nonreciprocal absorption and emission spectra are compared in **Fig. 3C** with apparent anti-symmetric properties despite nonreciprocity. For this case, $\tilde{P}_{cool}$ is only limited to 61.6 W/m$^2$ regardless of polarized selection, still being lower than the general reciprocal SE (in **Fig. 1B**). Therefore, merely breaking Lorentz reciprocity cannot improve the cooling efficiency and, in turn, will certainly weaken the radiative cooling performance (only the green switch is turned on in **Fig. 2B**).

**Simultaneous breaking thermal reciprocity and geometric symmetry make it possible to improve RC performance.**

The symmetric constraint of $\alpha(\lambda,\text{-}\theta)=e(\lambda,\theta), e(\lambda,\text{-}\theta)=\alpha(\lambda,\theta),\ \ \theta\in[0,\pi/2]$ leads to the reduction of cooling efficiency. Next, we consider selective nonreciprocal thermal emitters in the AW *without geometric symmetry*, which will provide a new degree of freedom to shape nonreciprocal emission and absorption to get rid of the angular constraint $e(\lambda,\theta,\mathbf{B}) = \alpha(\lambda,-\theta,\mathbf{B})$. We still assume that $e_r = e(\lambda,\theta) = 1, \alpha_r = \alpha(\lambda,\theta) = 0, \theta \in [0,\pi/2]$ in **Fig. 3D**, then the value of $e_{lx} = e(\lambda,-\theta)$ and $\alpha_{lx} = \alpha(\lambda,-\theta)$ can be independently tuned along with the results of $\tilde{P}_{cool}(T = T_{amb} = 20^{\circ}\mathrm{C})$ (middle surface) in **Fig. 3B** under the MLW condition. Two planes (red and grey) denote the theoretical limits of $\tilde{P}_{cool,SE}$ and $\tilde{P}_{cool,BE}$, both of which intersect with the surface with red and black crossing lines obtained. In this case, one can only realize cooling power enhancement by carefully choosing nonreciprocal and asymmetric cases above the intersecting lines, indicating that turning on both the green switch and red switch in **Fig. 1D** makes it possible to lighten the "*RC Enhancement*" bulb. But, to some extent, nonreciprocal and asymmetric thermal emitters in AW are not the most ideal candidates for nonreciprocal RC. **Figure 3E** compares the theoretical limit for different emitters marked in **Fig. 3B** with no conductive and convective heat exchange. The crossing points of these efficiency lines with $T = -273^{\circ}C$ and $T = 20^{\circ}C$ denote the value of atmospheric absorption loss and cooling power, respectively. In fact, the parasitic loss within AW is low, and the advantage of nonreciprocity in AW is not apparent. In contrast, the portion of the loss in non-AW is very large $P_{amb}(4 \sim 30\mu\mathrm{m}) - P_{amb}(4 \sim 13\mu\mathrm{m}) = 176\mathrm{W/m}^2$, but which has been long ignored.

**The role of non-atmospheric windows in nonreciprocal RC.**

The conventional wisdom of radiative cooling pays much attention to AWs, but we would like to emphasize that nonreciprocal RC will challenge this mind. Within the waveband of 13~30μm, the atmospheric transmission is considerably low for the MLW condition, which is responsible for the majority of parasitic loss. In **Fig. 3F**, we consider a thermal emitter with nonreciprocal and asymmetric spectra showing $e(13-30\mu\mathrm{m}, 0\sim\pi/2) = 1$ , $\alpha(13-30\mu\mathrm{m}, 0\sim\pi/2) = 0$ and $e(8-13\mu\mathrm{m}, -\pi/2\sim\pi/2) = \alpha(8-13\mu\mathrm{m}, -\pi/2\sim\pi/2) = 1$ . The calculated $\tilde{P}_{cool}(T_{amb} = 20^{\circ}\mathrm{C})$ (top plane) as functions of $\alpha_{lx} = \alpha(13-30\mu\mathrm{m}, -\pi/2\sim 0)$ and $e_{lx} = e(13-30\mu\mathrm{m}, -\pi/2\sim 0)$ are shown in **Fig. 3B** (top surface), totally suppressing all the cases working in AW. The grey star denotes the most ideal candidate whose emissivity is unitary among the whole angle and waveband but with zero absorptivity in 13~30μm. In this case, the majority of energy loss $P_{amb}$ can be efficiently recycled with a largely enhanced $\tilde{P}_{cool} = 274.2\text{W/m}^2$ (grey star), nearly three times larger than the reciprocal SE (red star). The $\tilde{P}_{cool}(T)$ of grey star under MLW condition will also significantly overpass that of the reciprocal BE under the ATRAN condition in **Fig. 1C**. It means that when considering breaking the efficiency limit of reciprocal radiative coolers, designing nonreciprocal and asymmetric thermal emitters at non-AWs could be an efficient solution. Further turning on the yellow switch in **Fig. 2B** makes the light bulb brighter. The inset in **Fig. 3E** compared the cooling temperature of different emitters mentioned above when considering convective influence ( $h = 8\text{W}\cdot\text{m}^{-2}\cdot\text{K}^{-1}$ ), in which nonreciprocal and asymmetric designs still show superior temperature reduction. Also, we give a proof-of-concept nonreciprocal nanostructure that can break Lorentz reciprocity and geometric symmetry simultaneously, composed of trial asymmetric multilayer nanobars in one period. Each nanobar has sixteen InAs layers with gradient doping concentrations from $Ne_1(\text{top}) = 5\times10^{17}\text{cm}^{-3}$ to $Ne_{14} = 18\times10^{17}\text{cm}^{-3}$ with $\Delta Ne = 1\times10^{17}\text{cm}^{-3}$ , $Ne_{15} = 20\times10^{17}\text{cm}^{-3}, Ne_{15} = 30\times10^{17}\text{cm}^{-3}$ (Details can be found in Supplementary Note 2). The period and width of nanobars are $p$=13μm, $w_1$=2.2μm and $w_2$=3.5μm. Here we merely pay our attention to a narrow non-AW waveband of $13.5\mu\text{m} \sim 16.5\mu\text{m}$. The asymmetric and nonreciprocal emission and absorption spectra are shown in **Fig. 3I**, along with typical spectra in $\theta = \pm30^\circ$ (**Fig. 3G**) under TM polarization. The cooling power *within this waveband* can be improved to 1.42 W/m$^2$ regardless of polarization selection, while the obtained value for reciprocal perfect BE is only 0.39 W/m$^2$. With these ideas in mind, further efforts can be devoted to considering how to broaden the nonreciprocal spectra in non-AW wavebands, relieve the influence of polarization-sensitive nonreciprocity, and further enhance the nonreciprocity between angular emissivity and absorptivity. Then, one could expect better cooling performance by introducing the nonreciprocal strategies into RC emitters design despite active methods of assistance.

**Discussion and Outlook**

In summary, under certain conditions, nonreciprocal RC could provide a new way to further improve RC performance, where one of the necessary things is to design structures with broken Lorentz reciprocity and geometric asymmetry simultaneously. Importantly, nonreciprocal and asymmetric thermal emitters could be game changers when facing terrible atmospheric conditions. The conventional RC suffers from high parasitic loss by the ambient environment, i.e., absorption, scattering, and reflection in non-AW waveband. Such ineffectiveness will be more and more prominent in an environment with high atmospheric humidity, high pollution, cloudiness, etc. These problems can be possibly solved using non-AW nonreciprocal ideas discussed here. In a word, conventional RC aims to make use of high transmission of the ambient environment, while nonreciprocal RC combined with asymmetric structures could be a game changer to make use of both atmospheric transmission and absorption. The cooling power in different atmospheric condition of various thermal emitters mentioned in **Fig. 3** are also compared in supplementary note 3. On the other hand, to be honest, designing nonreciprocal and asymmetric thermal emitters for radiative cooling remains challenging, especially considering broad mid-infrared wavebands. Besides, it should note that nonreciprocal RC will not be passive because of external energy input to break reciprocity. How to release the strict requirement of external stimulus is still an open question even in the area of nonreciprocal thermal photonics [8]. However, we remain believe that interdisciplinary collaboration between the areas of thermal nonreciprocity and RC will drive development and progress in nonreciprocal RC and make it come true in the near future.

## Acknowledgment

This work was supported by the National Natural Science Foundation of China (nos., 523060136, 52120105009, 52276078 and 52090063), Science and Technology Commission of Shanghai Municipality (Grants Nos.22ZR1432900) and China Postdoctoral Science Foundation (nos. BX20220200 and 2023M732199). C.-W.Q. acknowledges financial support from the Ministry of Education, Republic of Singapore (grant no. A-8000107-01-00), and the National Research Foundation, Singapore (NRF), under NRF's Medium Sized Centre: Singapore Hybrid-Integrated Next-Generation μElectronics (SHINE) Centre funding programme. The authors also thank the support from the Shanghai Jiao Tong University 2030 Initiative.

## Author Contributions

C.W.Q. and M.Q. L. conceived the ideas. M.Q.L. and S.H.J. performed the simulations. M.Q.L., S.H.J., C.L.Z. and B.X.W. analyzed the data. M.Q.L. wrote the paper with inputs and comments from all authors. C.W.Q. and C.Y.Z. supervised the project.

## Competing interests

The authors declare no competing interests.